\documentclass{iau}
\usepackage{graphicx}

\title[Star-formation in nuclear clusters] %% give here short title %%
{Star-formation in nuclear clusters and the origin  of the Galactic Center apparent core distribution}

\author{Danor Aharon \& Hagai B. Perets}

\author[Danor Aharon \& Hagai B Perets]   %% give here short author list %%
{Danor Aharon$^1$
 \and Hagai B Perets$^1$}

\affiliation{$^1$Physics Department, Technion - Israel Institute of Technology, Haifa,\\ Israel 32000,  email: {\tt danor@tx.technion.ac.il}}

\pubyear{2015}
\volume{312}  %% insert here IAU Symposium No.
\pagerange{119--126}
\jname{Star Clusters and Black Holes in Galaxies Across Cosmic Time}
\editors{A.C. Editor, B.D. Editor \& C.E. Editor, eds.}
\begin{document}

\maketitle

\begin{abstract}
Nuclear stellar cluster (NSCs) are known to exist around massive black
holes (MBHs) in galactic nuclei. Two formation scenarios were suggested
for their origin: Build-up of NSCs and Continuous in-situ star-formation.
 Here we study the effects of star 
formation on the build-up of NSCs and
its implications for their long term evolution and their resulting
structure. We show that continuous star-formation can lead to the build-up 
of an NSC with properties similar to those of the Milky-way NSC. We also find that the general
structure of the old stellar population in the NSC with in-situ star-formation
could be very similar to the steady-state Bahcall-Wolf cuspy structure.
However, its younger stellar population do not yet achieve a steady
state. In particular, formed/evolved NSCs with in-situ star-formation
contain differential age-segregated stellar populations which are
not yet fully mixed. Younger stellar populations formed in the outer
regions of the NSC have a cuspy structure towards the NSC outskirts,
while showing a core-like distribution inwards; with younger populations
having larger core sizes. 
\end{abstract}

\firstsection % if your document starts with a section,
\section{Introduction}

Nuclear stellar clusters, (NSCs) hosting massive black holes (MBHs)
are thought to exist in a significant fraction of all galactic nuclei.
Their origin is still not well understood. Two main scenarios were
suggested for their origin: (1) The cluster infall scenario, in which
stellar clusters inspiral to the galactic nucleus, disrupted, and
thereby build up the nuclear cluster (\cite{1975ApJ...196..407T,2013ApJ...763...62A}).
(2) The nuclear star formation (SF) scenario, in which gas infalls
into the nucleus and then transforms into stars through star formation
processes (\cite{1982A&A...105..342L}). Here we focus on the latter
process, and study the long term effects of SF on the formation and
evolution of NSCs. 

The structure, evolution and dynamics of NSCs have been extensively
studied in recent years. These explored the general dynamics of NSCs,
and in particular NSCs similar to the well-observed NSC in the Milky
Way Galactic Center (GC). The presence of a young stellar disk in
the central pc of GC, as well as dense concentration of HII regions
and young stars throughout the central 100 pc of the Milky-way (\cite{2004ApJ...601..319F})
provide evidence for a continuous star-formation in this region (\cite{2010RvMP...82.3121G}). Evidence for star
formation exists in other extagalactic NSCs (e.g.  \cite{2006AJ....132.2539S}).
\cite{2005ApJ...618..237W} argued that NSCs are protobulges that
grow by repeated accretion of gas and subsequent star formation, \cite{2006ApJ...650L..37M}
suggested a NSC in-situ star formation model regulated by momentum
feedback. 

These various studies provide further motivation and suggest that
star-formation has an important role in shaping NSCs and their evolution.
Here, we summarize our results of the role of in-situ SF in NSCs, and explore its
implication both for the build-up of the NSC, as well as the long
term evolution and structure of NSCs (detailed information can be found in \cite{2014arXiv1409.5121A}).
Our work makes use of the Fokker-Planck (FP) diffusion equations, first used 
by \cite{1976ApJ...209..214B} in this context, to describe the dynamics of 
stellar populations in dense clusters around MBHs.

\section{Evolution of NSCs around MBHs - by Fokker-Planck analysis}

NSCs are complex interacting systems. Their evolution and dynamics
are mainly affected by 2-body relaxation. Here we follow
the study the evolution of NSCs by numerically solving the FP equations
following the approach first used by Bahcall \& Wolf (BW;1976,1977)
in this context. However, we supplement the basic equations, for the
first time, with a source term accounting for SF, as well as use a
large number of distinct stellar populations to account for different
SF epochs. 

In our model we simulate the star formation in the GC through adding
an extra source term component to the FP equation. Its
value and range are determined according to the number of new stars
added in the appropriate region. We simulate multiple stellar populations
forming at different epochs, and follow the evolution of their distribution.
The modified FP equation with the addition of the source
term has the form:

\begin{equation}
\frac{\partial f(E,t)}{\partial t}=-AE^{-\frac{5}{2}}\frac{\partial F}{\partial E}-F_{LC}(E,T)+F_{SF}
\end{equation}

where The $F_{LC}(E,T)$ term corresponds to the empty loss-cone term 
(see \cite{1977ApJ...211..244L,1977ApJ...215...36Y,2007ApJ...656..709P}). 

There source term added to the FP equation is:

\begin{equation}
F_{SF}=\frac{\partial}{\partial t}\left(\Pi(E)E_{0}E^{\alpha}\right)\label{eq:source term}
\end{equation}

$\Pi(E)$ is a rectangular function, which boundaries correspond to
the region where new stars are assumed to from; $E_{0}$ is the source
term amplitude; and $F_{SF}$ is a power-law function with a slope
$\alpha$, defining the SF distribution in phase space . We simulated
a number of NSC evolutionary scenarios, taking different models for
the SF function (rate and spatial structure) and for the background
population. The chosen slope of the
SF function was motivated by the observed power-law (\cite{2009ApJ...703.1323D})
distribution of young stars observed in the young stellar disk in
the GC.

\section{RESULTS}

\begin{figure}[b]
\includegraphics[bb=10bp 18bp 943bp 413bp,scale=0.36]{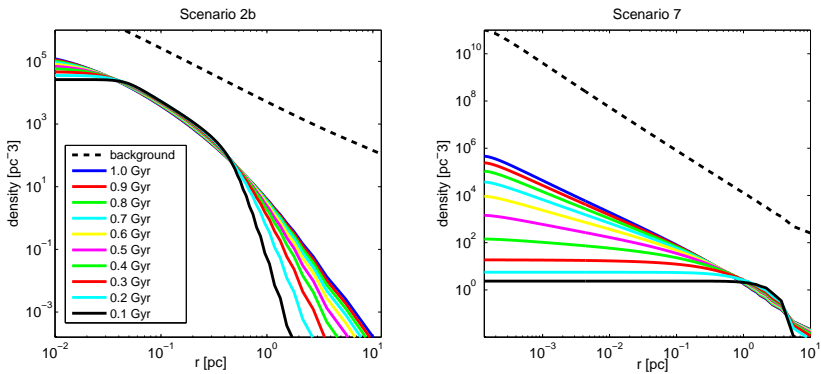}\protect\caption{\label{fig:profile}The number density profile of a 10 Gyrs evolved
NSC with SF inside the central 2 pc (scenario 2b) and outside (Scenario
7). The number of Gyrs presents the age of the new population. The
difference in the SF range affects the final distribution of an evolved
NSC. The old background stellar population (black dashed lines) correspond
either to stars produced through in-situ SF (over the first 9 Gyr;
right) or to a pre-existing BW cusp population (left). The structures
of these old populations show an almost BW-like steady state behaviour
in both of the models, while the young stellar populations formed
in the last Gyr are not yet relaxed, and show large cores ranging
in size. The young populations arise from in-situ star formation
at a rate of $10^{-4}\,{\rm stars\,}yr^{-1}$, during the last Gyr,
where each SF episode continues for 100 Myrs. }
\end{figure}

We have followed the evolution of multiple stellar populations formed
at different epochs. The build-up and structure evolution of an NSC
which grows through a continuous long-term in-situ star formation.
The final configuration of this NSC is very similar to that of a steady
state BW cusp, and the number densities are comparable to those observed
in the GC. The final structures of these NSCs after 10 Gyrs of evolution for the
two main scenarios (2b and 7) are summarized in Fig. \ref{fig:profile}.
The other scenarios can be found in \cite{2014arXiv1409.5121A}.
These models show the existence of a core-like structure for the young
stellar populations, where the cores vary in size, and are systematically
bigger for younger populations.

\section{DISCUSSION}

\subsection{The build up of nuclear stellar clusters through in-situ star formation\label{sub:The-build-up}}

We explored two evolutionary scenarios of NSCs: (1)
pre-existing NSCs with a BW-like structure that experience later SF
and (2) NSCs built-up completely from in-situ SF. Both type of models
assume that several epochs of gas-infall into the nuclear region triggered
SF, transforming the infalling gas into newly born stars. We show that NSCs built-up from in-situ SF give rise to NSCs dominated
by the stellar population formed at earlier stages (first few Gyrs).
The structure of the older population is very similar to a steady-state
BW-cusp, and the total number of stars is comparable to that inferred
for the GC NSC. We note that when lower mass stellar populations 
are assumed (e.g. if different initial mass function are considered)
the relaxation times become longer, as expected, and late-formed younger
populations are far from achieving a steady state structure, producing
larger core-like structures, as discussed in more details below.

\subsection{Core-cusp structure}

After a $\sim$10 Gyr of NSC evolution, the younger stellar
populations in NSCs may evolve to a core-like distribution, while
older population already become progressively cuspier. These results
are of great interest in light of the recent findings about the structure
of the GC NSC.

There are number of models suggesting to explain the origin of the GC core, 
for example \cite{2010ApJ...718..739M} suggested that a binary merger, or a
triaxial potential could deplete the inner regions of an NSC producing
a large core, and have shown that the long relaxation times would
not be sufficient to regrow a cusp. A similar behavior is seen in
our models, where progressively younger populations of stars formed
in the outer regions of the NSC do not relax and grow an inner cusp.
Though in both cases slow relaxation explains the non-growth of the
inner cusp, the origins of the initial core in both models differ,
and the outcomes could significantly differ as well. In particular,
the SF models studied here suggest that cores of different sizes could
exist for different stellar populations, and in particular an NSC
can have both a cusp distribution of old stars and a core distribution
for young and intermediate age stars.

\begin{figure*}[h]
\includegraphics[scale=0.38]{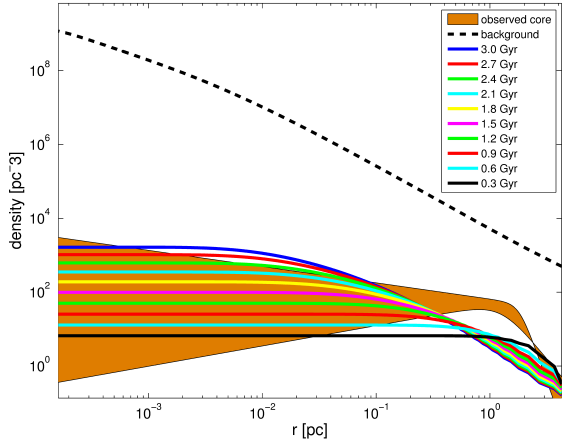}\protect\caption{\label{fig:core}The number density profile of the GC nuclear cluster
after 10 Gyr evolution (and total star formation epoch of 3 Gyr) compared
to the inferred range of number density profiles from 3D modeling
of GC observations \cite{2013ApJ...779L...6D}. The number of Gyrs
presents the age of the new population}
\label{obs vs theory}
\end{figure*}

We note that younger, more massive red giants could be more luminous
and more easily detected in observations (\cite{2011ApJ...741..108P}).
We therefore hypothesize that if such younger red-giants (up to 2-3
Gyrs old) are overly represented in observations then the observed
core could be limited to these younger populations, while the underlying
population of older stars might still have a cusp distribution. This
can be well demonstrated both in Fig. \ref{fig:core} where the model
results are compared with the 3D modeled number density profile determined
by \cite{2013ApJ...779L...6D} based on observations. As can be seen,
in some models a large, pc size core of up to a few Gyrs old stellar
populations can exist. Such a core might be consistent with the density
profiles inferred from observations, while the old stellar population
preserves a typical BW-like cusp profile.


\begin{thebibliography}{}

\bibitem[Aharon \& Perets 2014]{2014arXiv1409.5121A}
{Aharon, D. \& Perets, H.B.} 2014
\textit{ArXiv e-prints}, 1409.5121




%\bibitem[Busso et al. 1999]{Busso_etal99}
%{Busso, M., Gallino, R., \& Wasserburg, G.J.} 1999, 
%\textit{ARAA}, 37, 239







\bibitem[Antonini 2013]{2013ApJ...763...62A}
{Antonini, F.} 2013,
\textit{ApJ}, 763, 62


\bibitem[Bahcall \& Wolf 1976]{1976ApJ...209..214B}
{Bahcall, J.N. \& Wolf, R.A.} 1976,
\textit{ApJ}, 209, 214-232


\bibitem[Bahcall \& Wolf 1977]{1977ApJ...216..883B}
{Bahcall, J.N. \& Wolf, R.A.} 1977,
\textit{ApJ}, 216, 883-907



\bibitem[Do et al. 2009]{2009ApJ...703.1323D}
{Do, T., Ghez, A.M., Morris, M.R., Lu, J.R., Matthews, K., Yelda, S., \& Larkin, J.} 2009,
\textit{ApJ}, 703, 1323-1337


\bibitem[Do et al. 2013]{2013ApJ...779L...6D}
{Do, T., Martinez, G.D., Yelda, S., Ghez, A., Bullock, J., Kaplinghat, M., Lu, J.R., Peter, A.H.G., \& Phifer, K.} 2013,
\textit{ApJL}, 779, L6


\bibitem[Figer et al. 2004]{2004ApJ...601..319F}
{Figer, D.F., Rich, R.M., Kim, S.S., Morris, M., \& Serabyn, E.} 2004,
\textit{ApJ}, 601, 319-339




\bibitem[Genzel et al. 2010]{2010RvMP...82.3121G}
{Genzel, R., Eisenhauer, F., \& Gillessen, S.} 2010,
\textit{Reviews of Modern Physics}, 82, 3121-3195


\bibitem[Lightman \& Shapiro 1977]{1977ApJ...211..244L}
{Lightman, A.P. \& Shapiro, S.L. } 1977
\textit{ApJ}, 211, 244-262

\bibitem[Loose et al. 1982]{1982A&A...105..342L}
{Loose, H.H., Kruegel, E., \& Tutukov, A.} 1982,
\textit{AAP}, 105, 342-350



\bibitem[McLaughlin et al. 2006]{2006ApJ...650L..37M}
{McLaughlin, D.E., King, A.R., \& Nayakshin, S.} 2006,
\textit{ApJL}, 650, L37-L40


\bibitem[Merritt 2010]{2010ApJ...718..739M}
{Merritt, D} 2010,
\textit{ApJ}, 718, 739-761 



\bibitem[Perets et al. 2007]{2007ApJ...656..709P}
{Perets, H.B., Hopman, C., \& Alexander, T. } 2007,
\textit{ApJ}, 656, 709-720

\bibitem[Pfuhl et al. 2011]{2011ApJ...741..108P}
{Pfuhl, O., Fritz, T.K., Zilka, M., Maness, H., Eisenhauer, F., Genzel, R., Gillessen, S., Ott, T., Dodds-Eden, K., \& Sternberg, A. } 2011,
\textit{ApJ}, 741, 108 


\bibitem[Seth et al. 2006]{2006AJ....132.2539S}
{Seth, A.C., Dalcanton, J.J., Hodge, P.W., \& Debattista, V.P.} 2006, 
\textit{AJ}, 132, 2539-2555



\bibitem[Tremaine et al. 1975]{1975ApJ...196..407T}
{Tremaine, S.D., Ostriker, J.P., \& Spitzer, L. Jr.} 1975
\textit{ApJ}, 196, 407-411 

\bibitem[Walcher et al. 2005]{2005ApJ...618..237W}
{Walcher, C.J., Van der Marel, R.P., McLaughlin, D., Rix, H.W., B{\"o}ker, T., H{\"a}ring, N., Ho, L.C., Sarzi, M., \& Shields, J.C.} 2005,
\textit{ApJ}, 618, 237-246

\bibitem[Young 1977]{1977ApJ...215...36Y}
{Young, P.J.} 1977,
\textit{ApJ}, 215, 36-52


\end{thebibliography}
\end{document}